\documentclass[aps,prl,twocolumn,amsmath,amssymb,superscriptaddress,reprint]{revtex4-1}
\usepackage{amsmath}
\usepackage{graphicx}
\usepackage[dvipsnames]{color}
\usepackage{hyperref}
\usepackage{psfrag}
\usepackage{stmaryrd}
\usepackage{amssymb}
\usepackage{wasysym}
\usepackage{float}
\usepackage{color}

\usepackage[T1]{fontenc}
\usepackage[french,english]{babel}
\usepackage{lmodern}
\usepackage{graphicx} 
\usepackage{times}
\usepackage{amsmath,amssymb}
\usepackage{color}

\newcommand{\bnabla}{\boldsymbol{\nabla}}

\begin{document}

\title{Numerical validation of the inverse cascade of surface gravity wave action}

\author{Christopher Higgins}
\affiliation{Universit\'e Paris-Saclay, CNRS, CEA, Service de Physique de l'Etat Condens\'e, 91191 Gif-sur-Yvette, France.}
\author{Basile Gallet}
\affiliation{Universit\'e Paris-Saclay, CNRS, CEA, Service de Physique de l'Etat Condens\'e, 91191 Gif-sur-Yvette, France.}

\begin{abstract}
We report numerical simulations of surface gravity waves forced at small scale and the subsequent inverse cascade of wave action. We combine the spectral approach to simulating weakly nonlinear waves with the capabilities of modern Graphics Processing Units to reach unprecedented scale separation between the forcing and domain scales. The resulting broad inertial range allows for an unambiguous confirmation of the theoretical prediction for the spectrum in the inverse cascade regime, both in terms of spectral index and dependence of the spectral level on the action flux.
\end{abstract}

\maketitle

\textit{Introduction --.}  
Wave turbulence shares many similarities with standard hydrodynamic turbulence, including the transfer of conserved quantities in spectral space through self-similar cascades. As compared to hydrodynamic turbulence, however, an appealing aspect of Wave Turbulence Theory (also known as Weak Turbulence Theory, WTT in the following) is that it comes with a natural closure based on the dispersive nature of the waves and the timescale separation between linear and nonlinear processes, allowing for precise predictions for the energy spectrum \citep{nazarenko_book,zlf_book,galtier_book}. To wit, WTT begins with a weak-nonlinearity expansion, from which one derives a Wave Kinetic Equation (WKE) describing the slow evolution of the wave spectrum. In the inertial range of a turbulent cascade -- that is, over the range of scales on which forcing and dissipative processes do not directly act -- V.E. Zakharov first showed how one can derive exact self-similar solutions to the WKE. WTT has since been applied to many wave systems encountered in physics, providing a common framework for the description of out-of-equilibrium nonlinear dispersive wave systems: deep-water surface gravity and capillary waves \cite{falcon_et_al_2007,nazarenko_et_al_2010,deike_2014,campagne_et_al_2018,falcon_mordant_2022,zhang_pan_2022}, elastic waves on thin plates \cite{boudaoud_et_al_2008,humbert_2013,hassaini_et_al_2019,miquel_mordant_2011,mordant_2008}, inertial waves in rotating tanks \cite{yarom_sharon_2014,le_reun_2017,monsalve_2020}, internal waves in density-stratified fluids \cite{davis_et_al_2020,rodda_et_al_2022}, Bose--Einstein condensates \cite{zhu_et_al_2023,dyachenko_newell_pushkarev_zakharov_1992,nazarenko_onorato_proment_2014}, particle interactions \cite{gallet_nazarenko_dubrulle_2015} and gravitational waves \cite{galtier_laurie_nazarenko_2020}, to name a few.

Recently, there has been growing interest in understanding under which conditions the theoretically predicted turbulent states are realised in laboratory experiments and numerical simulations. The majority of these investigations are concerned with forward (direct) cascades, where the invariant -- often the energy -- is transferred from the injection scale through smaller scales, all the way down to dissipation. For systems that possess multiple quadratic invariants, however, wave turbulence predicts that some invariants may be transferred upscale through an inverse cascade mechanism. Deep-water surface gravity waves constitute one such system, conserving both energy and wave action. As for standard slowly-evolving waves (see e.g. waves in inhomogeneous media \cite{whitham_1974,buhler_2014}), wave action density is defined as wave energy density over the wave frequency. That is, with wavenumber $k$ and energy spectrum $E(k)$ , the wave action spectral density is $E(k)/\Omega_{k}$ where $\Omega_{k}=\sqrt{gk}$ is the angular frequency of surface gravity waves, with $g$ the acceleration due to gravity. In the absence of forcing and dissipation, the weakly nonlinear system conserves both the total wave energy, $\int_{0}^{\infty}E(k)dk$, and total wave action, $\int_{0}^{\infty}E(k)/\Omega_{k} dk$. When some forcing mechanism provides a source of energy and action, WTT applied to deep-water surface gravity waves predicts both a direct cascade of energy and an inverse cascade of wave action \cite{nazarenko_book,zlf_book}, the latter being characterised by the one-dimensional energy spectrum \cite{zakharov_zaslavskii_1982}:
\begin{equation}
E(k) = C_\text{KZ} \, g^{2/3} \zeta^{1/3} k^{-7/3}\, , \label{KZspectrum}
\end{equation}
where $\zeta$ denotes the wave action flux and the dimensionless prefactor $C_{\mathrm{KZ}}$ is the Kolmogorov--Zakharov (KZ) constant. Deep-water surface gravity waves can be considered the archetypal wave system for which an inverse cascade has been predicted. Additionally, the inverse cascade is thought to play a crucial role in the observed frequency downshift of wind-wave spectra, and in the formation of swell \cite{zakharov_2010}. Equally, a clear numerical or experimental validation of (\ref{KZspectrum}) would have consequences extending beyond surface gravity waves, to other systems for which inverse cascades have been predicted based on WTT: Bose-Einstein condensates and ‘optical turbulence’ as described by the Gross-Pitaevskii equation \cite{zhu_et_al_2023,dyachenko_newell_pushkarev_zakharov_1992,nazarenko_onorato_proment_2014,bortolozzo_2009}, Kelvin waves on thin vortex filaments \cite{kozik_svistunov_2005,nazarenko_2006,boffetta_2009}, spin waves \cite{lutovinov_chechetkin_1979,fujimoto_tsubota_2016} and most recently gravitational waves \cite{galtier_laurie_nazarenko_2020}.

Based on the literature, however, it appears that this inverse cascade is more difficult to observe than the forward cascade of surface gravity wave energy. Early experiments have shown reasonable agreement over a narrow inertial range \cite{deike_2011}, while more recent experiments in a much larger basin exhibit inverse transfers over approximately a factor of two in scale only \cite{falcon_et_al_2020}. While numerical simulations based on consistent truncations of the nonlinear surface wave equations are very successful at producing the expected direct cascade \cite{dyackenko_korotkevich_zakharov_2004,zhang_pan_2022}, numerical investigations of the inverse cascade again appear to be more challenging, such that the produced inertial range is either very narrow \cite{annenkov_shrira_2006}, or the observed spectral index departs from the theoretical prediction over a more extended inertial range \cite{korotkevich_2013}. That the inverse cascade is more challenging to observe is perhaps unsurprising: first, a broad inertial range requires forcing small-scale waves, while ensuring that these are unaffected by dissipation at yet-smaller scales. If the intuition gathered from idealised 1D systems holds for 2D, then the {\it frequency} inertial range must be broad, a particularly stringent criterion for the concave dispersion relation of deep-water surface gravity waves, as discussed in Ref. \cite{du_buhler2023}.
Secondly, the inverse cascade is very prone to finite-size effects, and in practice the first decade in $k$-space seems to exhibit features of discrete wave turbulence \cite{zhang_pan_2022b}. Such finite-size effects can lead to formation of a `condensate', inducing nonlocal transfers of action in spectral space and a spectrum that ultimately differs from the KZ prediction~(\ref{KZspectrum}) \citep{korotkevich_nazarenko_2023}. Thirdly, while the direct energy cascade has finite capacity and equilibrates over finite time, the inverse action cascade has infinite capacity: in an infinite domain, it would take an infinite amount of time to populate wavenumbers down to $k=0$ \cite{newell_rumpf_2011}. The practical consequence is that numerical simulations of the inverse cascade can take a very long wall-clock time (up to a year in the recent study reported in \cite{korotkevich_2023}!), with an inertial range that extends in an ever-slower fashion.

In this Letter we combine a higher-order spectral approach to simulating weakly nonlinear surface gravity waves \cite{west_etal_1987} with the capabilities of modern Graphics Processing Units to investigate the inverse cascade of wave action. This approach allows us to cope with the long integration times mentioned previously and to simulate waves over up to three decades in wavenumber (after de-aliasing). Such a broad range of scales makes it possible to force waves at sufficient distance from the small-scale dissipation, while simultaneously ensuring the existence of an inertial domain that does not involve the lowest decade of wavenumbers where discrete interactions dominate.

\begin{figure}
   \centerline{\includegraphics[width=9 cm]{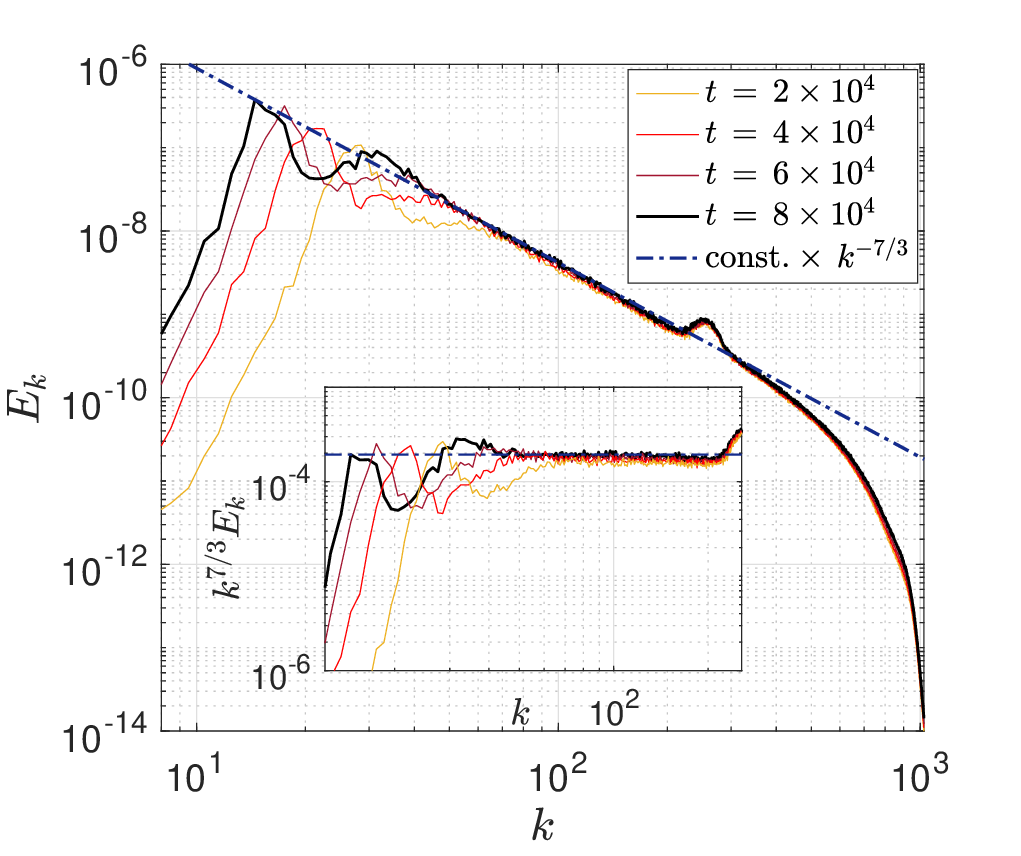} }
   \caption{Energy spectrum at successive times in a high-resolution simulation. The forcing ranges from $k_1=220$ to $k_2=292$. The dashed line indicates the theoretical value $-7/3$ for the spectral index. The rms surface slope is $0.14$. Inset: Spectra compensated by the theoretical power-law $k^{-7/3}$.
    \label{fig:longrun}}
\end{figure}

\textit{Numerical setup --.} We consider the evolution of gravity waves on the surface of an infinitely deep body of fluid. The problem domain is $(x,y)\in [0,2\pi L]^2$  with periodic boundary conditions. While the equations governing surface waves of arbitrary amplitude are remarkably challenging, restricting attention to the weakly nonlinear regime arising for weak wave slope allows for crucial simplifications. Specifically, retaining nonlinearities up to cubic order, the evolution of the wave field is governed by the following coupled PDEs for the free surface elevation $\eta(x,y,t)$ and the velocity potential evaluated on the free surface, $\psi(x,y,t)$:
\begin{eqnarray}
  \nonumber  & & \partial_t\eta = {D}\psi-\bnabla \cdot \left(\eta\bnabla\psi\right)-{D}\left(\eta {D}\psi\right)  \\
     & & +D\left[\eta{D}\left(\eta{D}\psi\right)\right] +\frac{1}{2}{D}\left(\eta^2\bnabla^2\psi\right)+\frac{1}{2}\bnabla^2\left(\eta^2{D}\psi\right) , \label{gov_EN_real} \\
\nonumber  & &      \partial_t\psi = -g\eta-\frac{1}{2}\left[|\bnabla \psi|^2-\left({D}\psi\right)^2\right]-\left({D}\psi\right){D}\left(\eta{D}\psi\right)\\
   & & -\left(\eta{D}\psi\right)\bnabla^2\psi \, .    \label{gov_EP_real}
\end{eqnarray}
where $\bnabla=(\partial_x,\partial_y)$ and the operator $D$ is defined in terms of the two-dimensional Fourier transform ${\cal F}$ as $D \psi = {\cal F}^{-1} \{ k {\cal F} \{ \psi \}  \}$, with $k$ the norm of the wavevector ${\bf k}$. To some extent, \eqref{gov_EN_real} and \eqref{gov_EP_real} make up the simplest set of equations describing the weakly-nonlinear evolution of deep-water surface gravity waves in the 2D horizontal plane. The wavy motion is described as a potential flow and the nonlinearities stem from the standard kinematic and dynamic boundary conditions at the moving fluid surface \cite{whitham_1974}. Crucially, this set of equations captures the spectral energy and action transfers between waves with different frequencies and wavevectors. Simpler models have been derived for the study of weakly nonlinear (quasi-)unidirectional waves, albeit in the shallow-water limit, such as the Korteweg-de Vries equation (1D) or the Kadomtsev–Petviashvili equation (weakly 2D). However, these models are based on drastic approximations, rendering them ‘integrable systems’. This rules out any resonant interaction between waves of different frequencies and wavevectors which are essential for wave turbulence to occur, and therefore also the associated spectral transfers of energy and action which are the focus of the present study \cite{zlf_book,zakharov_2009}.

In the following we nondimensionalise the equations using the lengthscale $L$ and timescale $\sqrt{L/g}$, keeping the same notations for the dimensionless variables. Decomposing $\eta(x,y,t)$ and $\psi(x,y,t)$ in terms of their Fourier amplitudes as 
$\left[ 
\eta(x,y,t) \, ,
\psi(x,y,t)
\right] = \sum_{{\bf k}\in \mathbb{Z}^2}  \left[ 
\hat{\eta}_{\bf k} (t) \, ,
\hat{\psi}_{\bf k} (t)
 \right] \, e^{i {\bf k}\cdot {\bf x}}$,
%
%
%\begin{eqnarray}
% \left[ \begin{matrix}
%\eta(x,y,t) \\
%\psi(x,y,t)
%\end{matrix} \right] = \sum_{{\bf k}\in \mathbb{Z}^2}  \left[ \begin{matrix}
%\hat{\eta}_{\bf k} (t) \\
%\hat{\psi}_{\bf k} (t)
%\end{matrix} \right] \, e^{i {\bf k}\cdot {\bf x}} \, , 
%\end{eqnarray}
%
we introduce the so-called `interaction variables':
\begin{eqnarray}
b_{\bf k}(t) = e^{i \Omega_k t} \left( \sqrt{\frac{\Omega_k}{2 k}}\hat{\eta}_{\bf k}  + i  \sqrt{\frac{k}{2 \Omega_k}}\hat{\psi}_{\bf k}   \right) \, .
\end{eqnarray}
The governing equations are recast in terms of $b_{\bf k}$ as:
\begin{equation}
    \partial_t b_{\bf k} = {\cal N}_{\bf k}+F_{\bf k}+\mathcal{D}_{\bf k}b_{\bf k},     \label{eqbk}
\end{equation}
where ${\cal N}_{\bf k}$ denotes the nonlinear terms and we have included a forcing term $F_{\bf k}$ and a damping term $\mathcal{D}_{\bf k}b_{\bf k}$. In the absence of forcing and dissipation, equations (\ref{gov_EN_real}-\ref{gov_EP_real}) -- and thus equation~(\ref{eqbk}) -- possess an {\it exact} energy invariant~\cite{dyackenko_korotkevich_zakharov_2004}:
\begin{eqnarray}
H  =  \frac{1}{2} & \iint_{[0,2\pi L]^2} & \left\{ g \eta^2 + \psi D \psi + \eta [|\bnabla \psi|^2 - (D \psi)^2] \right. \\
\nonumber & & \left. + \eta (D \psi) [D(\eta D \psi)+\eta \Delta \psi] \right\} \mathrm{d}x \mathrm{d}y \, .
\end{eqnarray}
Additionally, in the weakly nonlinear regime of interest here the wave action is an {\it adiabatic} invariant, conserved at the order of validity of the WKE (but not at the next order, see e.g. equation (5.2) of \cite{krasitskii_1994}). The wave action contained in scales larger than the forcing is defined as $A_\leftarrow(t)=\int_0^{k_1} \left< |b_{\bf k}|^2 \right>_\theta 2 \pi k \mathrm{d}k $, where $\left<\cdot\right>_{\theta}$ represents an average over the direction of the wavevector ${\bf k}$. The forcing term we choose is restricted to wavenumbers ${\bf k}=(k_x,k_y)$ satisfying the two conditions (i) $k_x > |k_y|$ and (ii) $k_1\leq k\leq k_2$, with the following complex amplitude:
\begin{equation}
F_{\bf k} = f_0 \frac{(k_2-k)(k_1-k)}{(k_2-k_1)^2}e^{i \chi_{\bf k}} \, . \label{eq:forcing}
\end{equation}
In this expression $f_0$ denotes the overall forcing amplitude, while $\chi_{\bf k}$ are time-independent random phases specified at the outset of a simulation. Directional localisation of the forcing -- condition (i) above -- has been observed to stabilise the numerical method and is therefore routinely used for long simulations with relatively strong forcing amplitudes \cite{onorato_2002,zhang_pan_2022,zakharov_2010}. More fundamentally, the relative spectral width of the forcing -- the ratio $(k_2-k_1)/(k_1+k_2)$ in condition (ii) above -- has recently been shown to have a crucial impact on the emergence of the turbulent cascade: a forcing which is too narrowbanded in wavenumber restricts the number of resonances partaking in the inverse cascade \citep{zhang_pan_2022b}, ultimately leading to a frozen form of discrete turbulence that halts the transfer of wave action to larger scales (see discussion section). 

Finally, following Refs.~\cite{onorato_2002,xiao_2013,zhang_pan_2022} the standard viscous dissipative operator is replaced by a high power of the Laplacian, $\mathcal{D}_{\bf k} = - \nu k^{30}$. Such hyperdiffusion behaves like a sharp low-pass filter~\cite{pushkarev_zakharov_2000} ensuring that only waves near the highest wavenumber allowed by the computational grid are directly subjected to damping. We have tested lower powers of the Laplacian and confirmed that the sharp dissipative operator does not result in an artificial bottleneck.

We solve equation (\ref{eqbk}) using a pseudospectral solver with 4$^{th}$-order Runge--Kutta time-stepping and de-aliasing following the $1/2$ rule. Graphical Processing Units are particularly well-suited for such low-memory simulations and lead to a significant speed-up, allowing us to perform long numerical simulations at high resolution.

\textit{Results --.} We first report a conservative numerical simulation where forcing is distant from the high-$k$ dissipative range and the broad inertial range does not include the first decade in spectral space where discrete turbulence may arise. To wit, we employ $4096^2$ resolution, which corresponds to $k_x$ and $k_y$ ranging from $-1024$ to $1024$ after de-aliasing, offering three decades in spectral space. The forcing ranges from $k_1=220$ to $k_2=292$, which proves broad enough to trigger an inverse action cascade. The waves remain weakly nonlinear throughout the entire simulation, the root-mean-square (rms) slope reaching approximately $\sqrt{\overline{\vert\bnabla\eta\vert^2}}=0.14$ at the end time of the simulation, where the overbar denotes the spatial mean. In Fig.~\ref{fig:longrun}, we plot the energy spectrum $E(k)=2 \pi k \Omega_k \left< |b_{\bf k}|^2 \right>_\theta$ at successive times. $E(k)$ is normalised such that its integral equals the quadratic part of the wave energy per unit surface, $\int_0^\infty E(k) \mathrm{d}k=\left< \eta^2/2  + \psi D \psi/2 \right>$, where angle brackets denote an average over several periods of the slowest waves and over the spatial directions. One can clearly see upscale transfers in Fig.~\ref{fig:longrun}, with a power-law spectrum developing over an increasingly broad inertial range as time goes on. To investigate compatibility with the KZ prediction (\ref{KZspectrum}), we provide an eyeguide with exponent $-7/3$ in the main figure and compensate the spectra by this prediction in the inset. The agreement is excellent over the inertial range, indicating that we have successfully circumvented any early-time frozen-cascade/condensate behavior at finite $k$.
\begin{figure}
    \centerline{\includegraphics[width=9 cm]{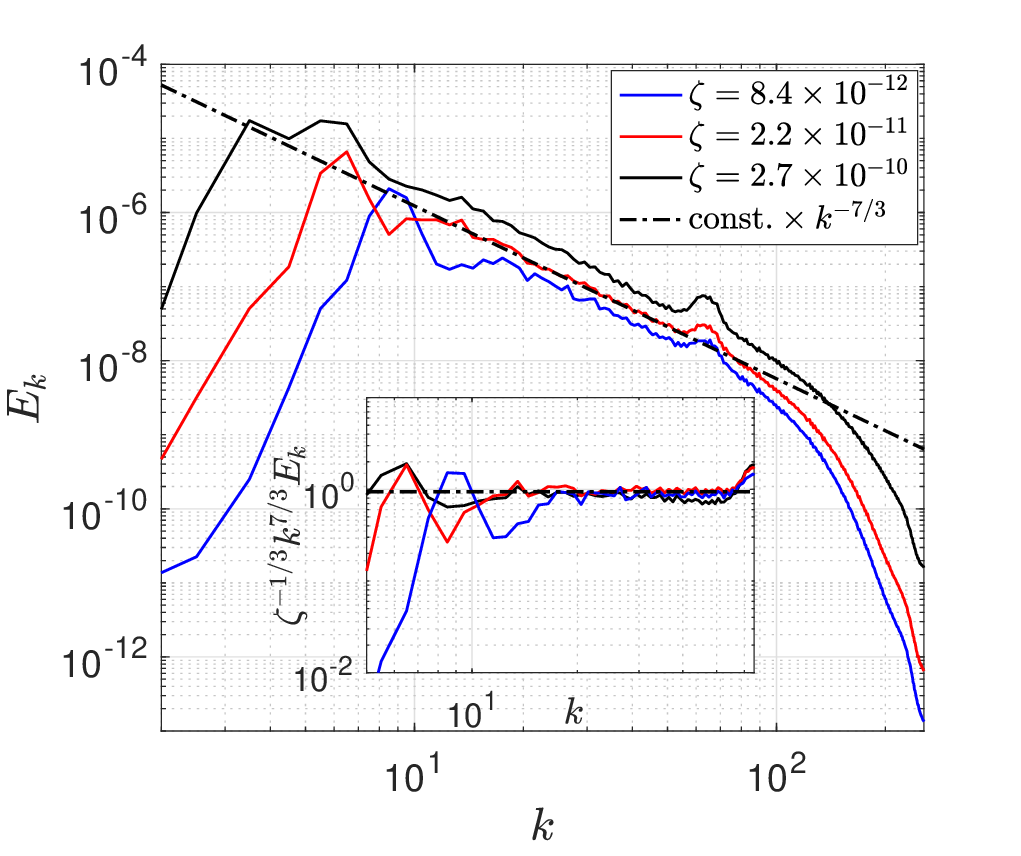} }
   \caption{Energy spectra at late time for various forcing amplitudes, see legend for corresponding values of the action flux $\zeta$. The rms slope of the fluid surface is $0.075$, $0.1$, and $0.15$ (bottom to top, respectively). The dashed line indicates the theoretical value $-7/3$ for the spectral index. Inset: Spectra compensated by the theoretical power-law $\zeta^{1/3} k^{-7/3}$. The collapse of the curves onto a plateau validates the theoretical prediction (\ref{KZspectrum}), the height of the plateau giving the value of the KZ constant $C_\text{KZ}$.
    \label{fig:shortruns}}
\end{figure}

It turns out that the inverse cascade can be reasonably observed with more modest resolution, and in Fig.~\ref{fig:shortruns} we show spectra obtained at $1024^2$ resolution for a maximum wavenumber of $256$ (after de-aliasing) and three different forcing amplitudes. Although the power laws are perhaps slightly less clean that in Fig.~\ref{fig:longrun}, they remain in excellent agreement with the spectral index of the theoretical spectrum~(\ref{KZspectrum}). A crucial ingredient explaining this success may be the forcing function~(\ref{eq:forcing}), which contributes to circumventing -- or delaying -- discrete turbulence processes in at least two ways. First, the relatively large spectral width $(k_2-k_1)/(k_2+k_1)\approx0.14$ allows for the activation of many resonances, and secondly the angular restriction permits stronger forcing amplitudes, thus inducing more nonlinear waves with broader frequency resonances. For each of the simulations in Fig.~\ref{fig:shortruns} we track the wave action $A_\leftarrow(t)=\int_0^{k_1} \left< |b_{\bf k}|^2 \right>_\theta 2 \pi k \mathrm{d}k = \int_{0}^{k_1}E(k)/\Omega_{k}dk $ contained in scales larger than the forcing scales, $k<k_1$. A time derivative leads to the instantaneous action flux $\zeta(t)=\mathrm{d} A_{\leftarrow}/ \mathrm{d}t$. 

We average $\zeta(t)$ over the last $10^4$ time units and use the resulting value to compensate the spectra in the inset of Fig.~\ref{fig:shortruns}. The collapse of the spectra onto a single constant curve validates the theoretical values for the exponents both in $\zeta$ and in $k$.

\begin{figure}
   \centerline{\includegraphics[width=9 cm]{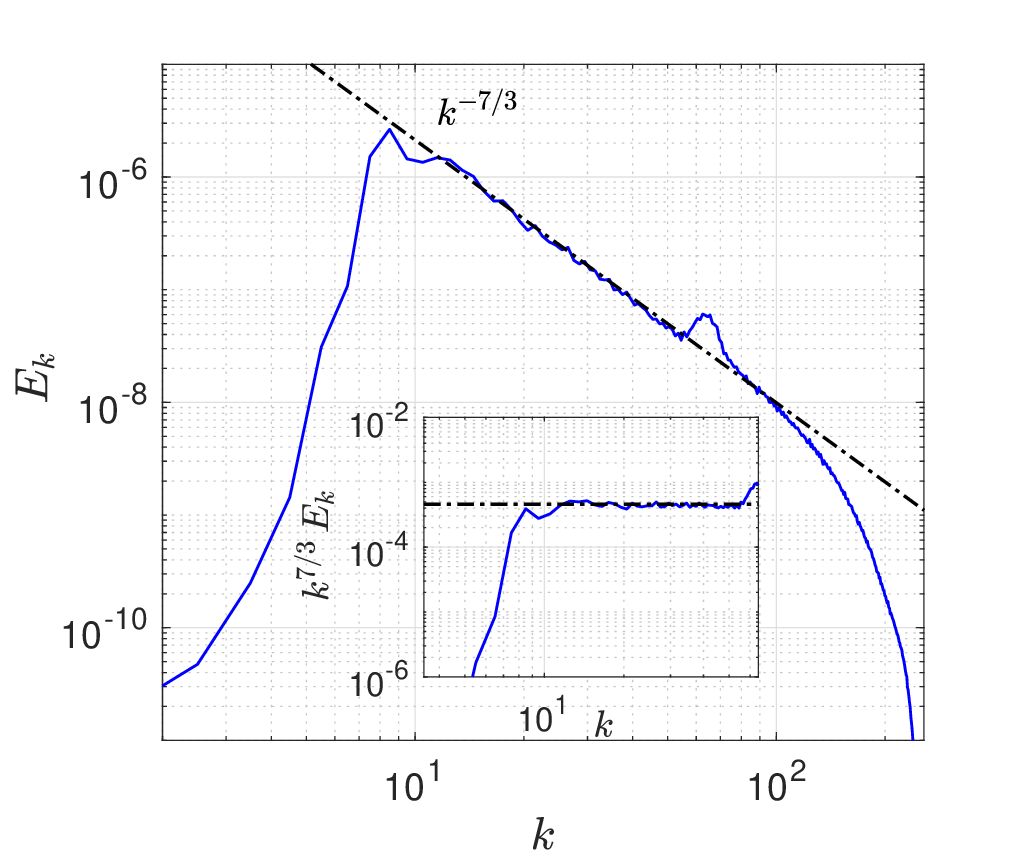} }
   \caption{Energy spectrum in statistically steady state for a run with the same forcing as the upper curve in Fig.~\ref{fig:shortruns}, with the addition of large-scale damping. The rms slope of the fluid surface is $0.13$. Once again, the spectral index agrees well with the theoretical prediction (\ref{KZspectrum}) as shown by the eyeguide in the main figure and the compensated spectrum in the inset. 
    \label{fig:mudamping}}
\end{figure}

\textit{Large-scale damping --.} One possible objection to both the present work and previous attempts reported in the literature is that the spectral index sometimes differs between the statistically steady state of a forced-dissipative turbulent cascade on the one hand, and the spectra observed during the transient phase where the forcing gradually populates the inertial range on the other hand \cite{connaughton_nazarenko_2004,connaughton_newell_2010,grebenev_2013,thalabard_2015}.  

To investigate this issue, we have performed an additional run with forcing similar to the upper curve in Fig.~\ref{fig:shortruns}, but with additional artificial large-scale damping in the form of an inverse Laplacian: $\mathcal{D}_{k} = -\mu k^{-2} - \nu k^{30}$. Such large-scale damping efficiently removes wave action at the end of the inverse cascade, leading to a stationary state. We plot the statistically-steady spectrum in Fig.~\ref{fig:mudamping}. The spectrum is similar to those in previous figures and the spectral index agrees equally well with the KZ prediction, as shown by the eyeguide in the main figure and the compensated plot in the inset.

\textit{Discussion --.} The present numerical simulations unambiguously validate the theoretical spectrum for the inverse action cascade of surface gravity waves, in terms of spectral index for both the wavenumber and the energy flux. A key ingredient of this success seems to be the choice of forcing function, which must be broad enough and strong enough to mitigate any freezing of the cascade at finite $k$. In this respect the present results strongly depart from previous attempts at observing the inverse cascade over an extended inertial range \cite{korotkevich_2012,korotkevich_2023}, where freezing of the cascade causes condensation at finite $k$. The resulting strong condensate then triggers nonlocal forward transfers of wave action associated with a distinct theoretical prediction for the spectral exponent \cite{korotkevich_nazarenko_2023}. 

Our observation of an extended inverse action cascade provides a unique opportunity to discuss the value of the associated KZ constant, estimated to be $C_\text{KZ} = 0.9 \pm 0.1$ based on the inset in Fig.~\ref{fig:shortruns} (our $4096^2$ run also agrees with this estimate). While the exact theoretical value is unknown, the estimate $C_\text{KZ}=2\pi \times 0.227\approx1.43$ has been proposed based on a nonlocal approximation in Ref.~\cite{zakharov_2010}, although a factor $1/2$ may be missing~\footnote{Pan \& Yue~\cite{pan_yue_2017} show that a factor $2$ is missed when angular-averaging the WKE in Zakharov's derivation of capillary wave turbulence. Correcting for this implies an extra $1/\sqrt{2}$ in the value of $C_{\mathrm{KZ}}$ predicted by \cite{pushkarev_zakharov_2000}. As surface gravity wave turbulence proceeds via four-wave interactions, two such angular averages are performed (e.g. \cite{dyachenko_newell_pushkarev_zakharov_1992}), hence a possible missing factor $1/2$}. The corrected theoretical estimate for $C_{\mathrm{KZ}}$ lies close to the numerical value obtained in this work. Beyond this approximate value it would be desirable to obtain the exact theoretical prediction for the action cascade KZ constant.

In the meantime, the careful study by Falcon et al.~\cite{falcon_et_al_2020} affords a comparison of the present results to experimental data. While their inertial range is arguably narrow, sufficient information is provided to extract an estimate of the KZ constant. The (dimensional) action flux is estimated as $\zeta=4\times 10^{-7}$m$^3$s$^{-2}$ and the elevation frequency spectrum takes the value $S_\eta(f) \simeq 2\times 10^{-4}$m$^2$s for $f=1.5$Hz. Recasting formula \eqref{KZspectrum} into the corresponding prediction for $S_{\eta}(f)$ and inserting the above quantities leads to: $C_\text{KZ}=2^{5/3} \pi^{8/3} S_\eta(f) f^{11/3} /(\zeta^{1/3} g) \simeq 0.8$. This rough estimate turns out to agree closely with our numerically determined value, and provides further evidence that Falcon et al. are in the right regime for development of an extended inverse cascade. That the cascade appears to saturate in their experiment may again be a consequence of the forcing mechanism, which is likely crucial at the experimental level too. It could be that a more spatially homogeneous and broadbanded forcing mechanism -- as opposed to boundary forcing -- would lead to a fully developed inverse cascade in such large-basin experiments.

Beyond surface gravity waves, it would be interesting to investigate to what degree the conclusions of the present study hold for other systems that have been predicted to sustain an inverse cascade based on WTT (as listed at the outset): is a broad enough forcing key to unfreezing the inverse cascade in these systems as well? Does this depend on whether the inverse cascade has finite or infinite capacity? How might this change with the number of waves involved in the dominant interaction mechanism? Careful experimental and numerical studies spanning a large variety of physical systems will be crucial to assess the extent to which WTT indeed provides a universal characterisation of out-of-equilibrium weakly nonlinear systems.

\section{Acknowledgements}
The authors thank Guillaume Michel for insightful discussions. We also thank the anonymous reviewers whose contributions and suggestions have helped to improve the clarity of our manuscript. This research is supported by the European Research Council under Grant Agreement FLAVE 757239.

\bibliography{InvCasc_arxiv}

\end{document}